\documentstyle[aasms4,12pt]{article}
\def\lax    {\ifmmode{_<\atop^{\sim}}\else{${_<\atop^{\sim}}$}\fi}
\def\gax    {\ifmmode{_>\atop^{\sim}}\else{${_>\atop^{\sim}}$}\fi}

\def\gtorder{\mathrel{\raise.3ex\hbox{$>$}\mkern-14mu
	     \lower0.6ex\hbox{$\sim$}}}
\def\ltorder{\mathrel{\raise.3ex\hbox{$<$}\mkern-14mu
	     \lower0.6ex\hbox{$\sim$}}}
\def\gsim{\mathrel{\raise.3ex\hbox{$>$}\mkern-14mu
	     \lower0.6ex\hbox{$\sim$}}}
\def\lsim{\mathrel{\raise.3ex\hbox{$<$}\mkern-14mu
	     \lower0.6ex\hbox{$\sim$}}}

\begin{document}

\title{Modeling The Time Variability of Accreting Compact Sources}

\author{Demosthenes Kazanas \& Xin-Min Hua\altaffilmark{1}} 
\affil{LHEA, NASA/GSFC Code 661, Greenbelt, MD 20771}

\altaffiltext{1}{Universities Space Research Association}
%\altaffiltext{2}{CSI, George Mason University}

%\vskip 0.5 truecm
%\font\rom=cmr10
%\centerline{\rom In press, Astrophys. J. (Letters)}

\begin{abstract}

We present model light curves for accreting Black Hole Candidates 
(BHC) based on a recently proposed model for their spectro-temporal 
properties. According to this model, the observed light curves and
aperiodic variability of BHC are due to a series of soft photon 
injections at random (Poisson) intervals near the compact object and 
their reprocessing into hard radiation in an extended but non-uniform 
hot plasma corona surrounding the compact object. We argue that the 
majority of the timing characteristics of these light curves are due  
to the stochastic nature of the Comptonization process in the 
extended corona, whose properties, most notably its radial density 
dependence, are imprinted in them. We compute the corresponding Power 
Spectral Densities (PSD), autocorrelation functions, time skewness of 
the light curves and time lags between the light curves of the sources 
at different photon energies and compare our results to observation. 
Our model light curves compare well with observations, providing good 
fits to their  overall morphology, as manifest by the autocorrelation 
and skewness functions. The lags and PSDs of the model light curves 
are also in good agreement with those observed (the model can even 
accommodate the presence of QPOs). Finally, while most of the 
variability power resides at time scales $\gsim$ a few seconds, at 
the same time, the model allows also for shots of a few msec in 
duration, in accordance with observation. We suggest that 
refinements of this type of model along with spectral and phase lag 
information can be used to probe the structure of this class of 
high energy sources.

\end{abstract}


\section{INTRODUCTION}

The study of the physics of accretion onto compact objects (neutron stars
and black holes) whether in galactic (X-ray binaries) or extragalactic
systems (Active Galactic Nuclei) involves length scales much too small to
be resolved by current technology or that of the foreseeable future. As such, 
this study is conducted mainly through the theoretical interpretation of 
spectral and temporal observations of these systems, much in the way
that the study of spectroscopic binaries has been used to deduce the 
properties of the binary system members and the elements of their
orbit. In this endeavor, the first line of attack in unfolding their 
physical properties is the analysis of their spectra. 
For this class of objects and in particular the Black Hole Candidate 
(BHC) sources, a multitude of observations have indicated 
that their energy spectra can be fitted very well by Comptonization of soft 
photons by hot electrons; the latter are ``naturally" expected to be 
present in these sources, a result of the dissipation of the accretion 
kinetic energy within an accretion disk. It is thus generally agreed upon that 
Comptonization is the process by which the high energy ($\gsim 2-100$  keV)
spectra of these sources are formed, with the study of this process receiving 
hence great attention over the past couple of decades  (see e.g. Sunyaev \& 
Titarchuk 1980; Titarchuk 1994; Hua \& Titarchuk 1995). Thus, while the issue 
of the detailed dynamics of accretion onto the compact object is still not 
resolved, the assumption of the presence of a thermal distribution of hot 
electrons in the vicinity of the compact object has proven sufficient to 
produce models which successfully fit the spectra of the emerging 
high energy radiation. Spectral fitting has subsequently been employed 
as a way to constrain or even determine the dynamics of accretion onto 
the compact object.  

It is well known, however, that the Comptonization spectra cannot 
provide, in and of themselves, any information about the size of the 
scattering plasma, because they depend (in the case of optically thin 
plasmas) on the product of the electron temperature and total probability 
of photon scattering in the plasma, a quantity proportional to its Thomson depth.
Therefore, they cannot provide any clues about the dynamics of accretion 
of the hot gas onto the compact object, which require the knowledge of the 
density and velocity as a function radius. To determine the dynamics of 
accretion, one needs, in addition to the spectra, time 
variability information. It is thought, however, that such information 
may not be terribly relevant, because it is generally accepted that the 
X-ray emission originates at the smallest radii of the accreting flow 
and as such, time variability would simply reflect the dynamical or 
scattering time scales of the emission region, of order of msec for 
galactic accreting sources and $10^5 -10^7$  times longer for AGN.
  
Recent RXTE (Focke, private communication) as well as older HEAO-1 observations 
(Meekins et al. 1984) of the BHC Cyg X-1, which  resolved X-ray flares of 
duration $\sim$ a few msec, appear to provide a validation of our simplest
expectations. At the same time, however, the X-ray fluctuation
Power Spectral Densities (PSD) of accreting compact sources generally 
contain most of their power at frequencies $\omega \lsim 1$ Hz, far removed 
from the kHz frequencies expected on the basis of the arguments given above.
Flares of a few msec in duration, while present in the X-ray light 
curves of Cyg X-1, contribute but a very small fraction to its overall
variability as manifest in its PSD, which exhibits very little power
at frequencies $\gsim$ 30 Hz (Cui et al. 1997). Interestingly, this form of 
the PSD has been reported in addition to Cyg X-1 also in the BHC GX 339-4 and
Nova Muscae GINGA data (see e.g. Miyamoto et al. 1991).

This discrepancy between the observed and the expected distribution of the 
variability power of BHC sources, hints that one may have to revise 
the notion that the entire hard X-ray emission of these sources derives 
from a region a few Schwarzschild radii in size and indicates the need of 
more detailed models of the timing properties of these systems. In this
respect, models of the timing properties of accreting compact sources
have been rather limited (with the exception of models of the quasi-periodic 
oscillations), the reason being, (on the theoretical side) the largely 
aperiodic character of their light curves and (on the experimental side) 
the lack of sufficiently large area detectors, in conjunction with high 
telemetry rates which would provide high timing resolution 
data. As a consequence, the study and modeling of these class of 
sources has concentrated mainly on their
spectra, whose S/N ratios can be improved by longer exposure times. 
Thus, the literature associated with models of the spectra of BHC sources 
is substanially larger than that of their timing models. 

Much of the earlier work in modeling the aperiodic light curves of BHC sources
was kinematic in character, aimed in its deconvolution in elementary
events with the goal of providing fits to the observed PSDs. Thus, Lochner 
(1989) was able to reproduce the observed PSDs as the ensemble of incoherent, 
exponential shots of durations ranging from 0.01 to 1 secs, while Lochner, 
Swank \& Szymkoviak (1991) searched (unsuccessfully) for a low dimensional 
attractor in the light curve of Cyg X-1. In a similar fashion, Abramowicz et 
al. (1991) produced model light curves with PSDs similar to those 
observed, resulting from a large number of bright spots in an accretion disk. 
More recently, a number of dynamical models have appeared in the literature:
Chen \& Taam (1995) produce time variability as a result of hydrodynamic
instabilities of an accretion disk, while Takeuchi, Mineshige \& Negoro (1995)
use a model of self-organized criticality to simulate accretion onto the compact
object. Both these models provide reasonable fits to the observed PSD shapes
(however, not necessarily to their normalizations) by producing a 
modulation of the accretion rate onto the compact object. 

Models of the type just described, generally aim to account for only
the simplest  of variability tests, namely the PSD, and usually they 
have enough freedom to be able to match the shape of the observed
PSDs. However, in most of them, the associated light curves are very much 
different than those observed, testimony to the fact that the power 
spectrum erases all the phase information available in the signal and 
that very different light curves can in fact have identical PSDs.
We are aware of only one effort to keep track and search for correlations
in the phases of the different Fourier components in the light curves 
of low luminosity AGN (Krolik, Done \& Madejski 1993). 
The conclusion of this search was that the process responsible for the
AGN variability (at least the ones studied in the above reference)
is incoherent, in the sense that the phases of the various Fourier
components are uncorrelated and apparently random.

In this respect, 
%With respect to the issue of the phases associated with the variability
%of these type of sources 
one should note that the very fact that the spectra  of these sources 
are due to Comptonization of soft photons has several direct implications
concerning the phases of photons of different energies:
Because, on the
average, it takes longer to produce a photon of high energy compared 
to that of a lower energy, the energy of the escaping photons increases 
with their residence time in the scattering medium. Thus, while there 
may not be any coherence in the {\it absolute} phases of the various 
Fourier components of the light curves of 
this class of sources, the {\it relative} phases between photons
in two different energy bands may not be random, provided that the observed 
radiation is produced by a single hot electron cloud rather than a large 
number of individual, disjoint sites. As a result, the 
hard photon light curves should lag with respect to those of softer photons 
by amounts which depend on the photon scattering time in the plasma. 
If the sources are
of rather limited spatial extent and the scattering takes place only
in the vicinity of the compact object, then the corresponding time lags
should be roughly constant (independent of the Fourier frequency) and 
of order of the scattering time in this medium, i.e. $\simeq 10^{-3}$ 
sec for galactic
sources and correspondingly larger for AGN. On the other hand, if the
scattering medium is extended, the lags should cover a 
temporal range which depends on the characteristics of the scattering
medium.

Searches for these lags in the light curves of the BHC sources Cyg X-1 
and GX 339-4 in the GINGA (Miyamoto et al. 1988, 1991) and the 
more recent RXTE data (Cui et al. 1997b), have detected their 
presence and, more important, discovered that the hard time lags
$\Delta t$ depend on the Fourier period $P$,  increasing roughly linearly with  
$P$ from $\Delta t \lsim $0.001 sec at $P \simeq 0.05$ sec to $\Delta t \lsim $
0.1 sec for $P \simeq 10$ sec. These long lags and in particular
their dependence on the Fourier period $P$ are very difficult to interpret
in the context of a model in which the X-ray emission is 
due to soft photon Comptonization in the vicinity of the compact object; 
in such a model, the hard lags should simply reflect the photon scattering time 
in the specific region (i.e. $\simeq$ msec) and moreover, they should be
independent of the Fourier frequency. While this type of time lag was
found originally in the light curve of the BHC Cyg X-1, similar lag behavior
has been recorded also for the transient J0422+32 (Grove et al. 1998) and
also the source GRS 1758-258 (Smith et al. 1997). Finally, to indicate that 
these results may not be universal, the data of Wilms et al. (1997) indicate
that the lags associated with Cyg X-1 during their observation, increase 
much more gradually. 

Motivated by the discrepancies between our expectations of X-ray
variability based on simple dynamical models and the observed form of  
their PSDs and their frequency dependent hard lags, we have revisited the
issue of time variability of BHC sources (Kazanas, Hua \& Titarchuk 1997, 
hearafter KHT; Hua, Kazanas \& Titarchuk 1997, hearafter HKT; Hua, 
Kazanas \& Cui 1999, hereafter HKC). 
The central point of our considerations has been that, 
contrary to the prevailing notions about the timing behavior of accreting
sources, features in the observed PSD correspond not to variations in the 
accretion rate but rather to properties of the electron density distribution 
of an extended ($\gsim 10^3 R_s$; rather than a compact  $\lsim 
10 R_s$) hot electron corona. With this assumption in place, features in 
the PSDs translate to well defined properties of the scattering corona
which can in principle be checked for consistency with observations.
Thus we have proposed that the low frequency ($\nu \lsim 1$ Hz) break
of the PSD to white noise is associated with the outer edge of the Comptonizing
corona,  while its inner edge is very close to the compact source. 
We further indicated that the power law - like PSDs depend on both the 
density profile and the total Thomson depth of the corona, implying 
correlations between the energy and the variability power spectra, 
which can be sought for in the data.
 
Using this extended corona model, we were able to show explicitly 
that the observed energy spectra convey rather limited information
about the structure of the scattering medium (KHT, HKC) 
and that very different hot electron configurations
can indeed yield the same energy spectrum as long as the {\it total} 
probability of scattering remains the same across the different configurations.

We  have also shown (HKC) that the fact that the 
lags between different energy bands depend on the {\it differential} 
probability of scattering at a given range of radii (contrary to the 
spectrum which depends on the total probability) affords a
means of {\it probing the density structure of the corona} through the
study of the lag dependence on the Fourier period $P$. Finally, we were 
able to show that due to the linearity  of Compton scattering, the 
coherence function (as defined by Vaughan \& Nowak 1996) of the 
extended hot corona configuration is very close to one, in agreement 
with observations (Vaughan \& Nowak 1996), suggesting, in addition, 
that the properties of the scattering hot electron cloud remain constant 
over the observation time scales (Hua, Kazanas \& Titarchuk 1997). 

Thus the timing observations indicate, on one hand, that the absolute 
phase of the light curves of accreting compact objects are random, 
implying an incoherent process, while on the other hand that the 
relative phases between different energy bands are indeed extremely well 
correlated, indicating a coherent underlying process. The puropose 
of the present paper is to produce model light curves of accreting 
compact sources compatible with these apparently contradictory aspects
of their coherence as well as the observed PSDs. We also examine the 
structure of the latter (including the presence of QPOs) and their 
dependence on the sources' luminosity in the general context of these 
models. Having provided a prescription for producing model light curves, 
we subsequently examine their properties in the time rather than 
frequency domain and indicate tests which may confirm or disprove the 
fundamentals of our model.

In \S 2 we outline the extended corona configuration under 
consideration and we present models of the Comptonization response 
function and the corresponding power spectra associated with it. 
We also elaborate on the fundamental tenet of our models, namely 
that of  association of PSD features  with corresponding features 
in the density structure of the extended corona by providing a generic
QPO model within this framework. 
In \S 3 we provide a prescription for generating model light curves 
using  the coronal response function and we do generate a number of 
such curves for different values of the model parameters. With the 
model light curves at hand, we further elaborate our analysis by 
computing and comparing their attributes to observation: Thus in 
\S 4 we compute phase lags as a function of the Fourier frequency 
and in \S 5 the associated autocorrelation and skewness functions. 
Finally, in \S 6 the results are summarized and discussed.

\section{THE EXTENDED CORONA AND ITS RESPONSE}

Kazanas, Hua \& Titarchuk (1997) proposed that the timing 
properties of BHC discussed above, i.e. the PSD and the form 
and frequency dependence of the 
time or phase lags, can be easily accounted for with the assumption that 
all associated variability is due to the Compton scattering of soft 
photons in an extended, non-uniform corona surrounding the compact 
object and spanning several decades in radius. Specifically, they 
proposed the following density profile for the Comptonizing medium:

$$n(r) = \cases  {n_1 &for $r \le r_1$ \cr n_1 (r_1/r)^{p} &
for $r_2 > r > r_1$ \cr} \eqno(1)$$

\noindent
where the power index $p >0$ is a free parameter; $r$ is the radial 
distance from the center of the spherical corona; $r_1$ and $r_2$ are 
its inner and outer radii respectively. 

KHT considered $p$ to be an arbitrary parameter, however, 
as indicated in that reference, the value of $p=1$ allows for scattering
of the photons with equal probability over the entire extent of the 
corona introducing time lags at every available frequency with equal 
probability, a fact which is in agreement with the observations; hence they
considered the value $p=1$ as the fiducial value of this parameter. 
However, they also  considered different values of $p$, in particular 
the value $p=3/2$, as it corresponds to the density profiles of the 
currently popular Advection Dominated Accretion Flows (ADAF; 
Narayan \& Yi 1994). 

\subsection{The Shot Profiles}

The time response of the given corona, i.e. the flux of photons escaping
at a given energy range as a function of time, following the injection 
of a $\delta$-function of soft photons at its center at $t=t_0$ has been 
computed using the Monte Carlo code of Hua (1997). As  discussed in KHT 
and in HKC its form, $g(t)$, ignoring its  rising 
part,  can be approximated by a Gamma distribution function, i.e. a 
function of the form
$$g(t) = \cases{t^{\alpha -1}e^{-t/\beta}, &if $t\ge 0$; \cr
                0 &otherwise, \cr } \eqno(2)$$

\noindent
where $t$ is the time; $\alpha > 0$ and $\beta > 0$ are parameters determining
the shape of the light curves which depend on the scattering depth $\tau_0$
and photon escape energy (see Figures 1a \& 1b in KHT). As indicated in KHT,
for small values of the total depth of the corona $\tau_0 \ll 1$, 
$\alpha -1 = -p$, while as $\tau_0$ increases the power law part of the curve
becomes progressively flatter, i.e. $\vert \alpha -1 \vert < p$. 
For the cloud with $p=1$ in Eq. (1), $\alpha$ is small compared to 1 and 
$\beta$, determined by the outer edge of the scattering cloud, is taken to 
be of the order of 1 second so that the light curves have exponential cutoff 
at $\sim 1$ second. For uniform cloud, $p=0$, $\alpha=1$ and the light curve
is a pure exponential (leading to a PSD proportional to $\omega^{-2}$); 
one should note though that the precise value of $\alpha$ depends also on
the photon energy). The simplified form of the response function given 
by equation (1) allows one to compute analytically its Fourier transform 
(HKC)
$$G(\omega) = \displaystyle{{\Gamma(\alpha) \beta^{\alpha}}\over
              {\sqrt{2\pi}}}(1+\beta^2\omega^2)^{-\alpha/2}
              e^{i\alpha\theta}. \eqno(3)$$

\noindent
where $\Gamma(x)$ is the Gamma function; $\omega$ is the Fourier frequency 
and $\theta$ is the phase angle and $\tan{\theta} = \beta\omega.$  

However, the form of the realistic response functions is more complicated.  
Figure 1 shows the shape of the response function as computed using the Monte 
Carlo code (Hua 1997).
The parameters used in the Monte Carlo calculation were, $r_1 = 6.35 \times 
10^{-3}$  light sec $=1.9 \times 10^8$ cm, $r_2= 6.24$ light sec$ = 1.87 
\times 10^{11}$ cm, the electron temperature was taken to be $T_e = 100$ keV, 
the Thomson depth $\tau_0 = 1$ and the density at bottom of the corona 
$n_1 = 10^{15}$ cm$^{-3}$. The soft photons were chosen randomly from a Planck
distribution of temperature 1 keV and the data give the flux of Comptonized 
photons in the $35 - 60$ keV energy range.

To simulate the precise form of the shots in the coronae we consider
we use a more complex function
$$g(t) = \cases{A_1 (1-B_1 x^b)x^{\gamma}, &if $x \equiv t/t_0 \le 1$; \cr
                A_2 (1+B_2x^{-b})x^{\alpha -1}e^{-(xt_0/\beta)^3}, 
&if $x \equiv t/t_0 > 1$, \cr } \eqno(4)$$

\noindent
where $b, \gamma > 0$. The parameters $A_2, B_1, B_2$ are given in terms
of the arbitrary normalization $A_1$ and the form parameters $\alpha, \beta,
\gamma, t_0$ and $b$ by the requirement that the function $g(t)$ and 
its first derivative be continuous at $x = 1$, with that point being also a local
maximum. The parameters $\alpha, \beta$ have the same meaning as those used
in Eq. (2) above, while $t_0$ and $\gamma$ indicate the time at which the 
response function achieves its maximum value (of order $r_1/c$) and the 
rate at which this maximum is achieved. The cut-off form of $g(t)$ is steeper 
than exponential to mimic the detailed curve produced by the Monte Carlo
simulation. One should note the additional parameter $b$, which regulates the
``sharpness" of the transition from the rising to the falling part of the 
model response function. The values of the parameters used in fitting the 
response function were $\alpha = 0.4, ~\gamma = 1.5$, $\beta = 10$ sec, 
$t_0 = 0.02$ sec, $b=1$.  In the same figure we also present the PSD which
corresponds to a single shot with this specific form of the response 
function. As will be argued in the next section, under certain assumptions,
 this is also the PSD  of the entire light curve.

The particular light curve and the associated values of the corresponding 
fits should be considered only as fiducial values. Monte Carlo runs 
with smaller values of $r_1$ gave shots achieving their peak flux at 
times scales proportionally shorter. Thus fits of corona response 
functions with $r_1 = 4.77 \times 10^{-4}$  light sec $= 1.43 \times 
10^7$ cm and $n_1 = 10^{16}$ cm$^{-3}$, gave rise times of order 
$t_0 = 0.001$ sec.

\subsection{The Power Spectra}

%In this section we present samples of the form of the response function of 
%the extended corona discussed above for different values of the parameters
%and also their associated PSD, to indicate that their form is in general 
%accordance with those observed, that, as far as this statistic of variability is %concerned, our model is in general agreement with observation. 

As will be discussed in the next section, the PSDs of the model X-ray 
light curves of accreting neutron stars and BHC sources which we present,
to a large extent, reflect the properties 
of the response functions of the corresponding coronae;
we therefore feel that a short discussion of their form is, at this point, 
necessary. To begin with, we note that in the limit of infinitely sharp 
turn-on of the shots under consideration, the Fourier transform of $g(t)$ 
is given by $G(\omega)$ of Equation (2) and therefore the PSD, under these 
conditions, can be computed analytically (HKC)
$$\vert G(\omega) \vert ^2 = {\Gamma(\alpha)^2 \beta^{2\alpha}\over
              2\pi}(1+\beta^2\omega^2)^{-\alpha} ~. \eqno(5) $$

As discussed in KHT, these PSDs consist of a power law section with slope 
which is related to the corresponding slope of the power law segment 
of the response function while flattening to white noise for $\omega \lsim 
1/\beta$. Since we have adopted the value $\alpha \simeq 0.5$ in the form of the 
response function, the PSD consists mainly of a power law 
segment of slope $2 \alpha -2 \simeq -1$, i.e. flicker noise, representative
of the power law segment of the time response function. 
However, for the case of more realistic shots which have a finite 
rise time, such as those shown in fig. 1, the computation of the 
PSD can no longer be done analytically. Moreover, the finite size of the 
shot rise time introduces additional structure in the PSD which manifests 
as a break in its high frequency portion. In figure 1 we present, 
in addition to the profiles of the response functions, the corresponding  
PSDs (short-dashed lines).

It is of interest to compare the shape of the PSD to those associated with 
the real data of the BHC Cyg X-1, GX 339-4 and Nova Muscae (Miyamoto et al. 
1992, fig. 1). One can see that, in addition to the flicker noise behavior, 
our model also produces the steepening observed at higher  frequencies 
(see also Cui et al. 1997 for RXTE data of Cyg X-1 and Grove et al. 1998
for OSSE data of GRO J0422+32). The similarity of 
the model PSD shapes to the data notwithstanding, the most important 
feature of the present model is, in our view, the physical association
between the specific PSD features and the physical characteristics
of the hot electron corona responsible for the production of the high 
energy radiation through Comptonization (most notably its size and 
radial density slope, though the latter is only determined unequivocally
through the hard lags, HKC). 

We would like to comment specifically on the effect of the parameter $b$, 
the shot ``sharpness"  parameter, on the form of the PSD; as one might 
expect, for values $b < 1$, this parameter can affect significantly the 
form of the PSD, rendering it steeper than expected 
on the basis of the value of the parameter $\alpha$. However, the realistic
light curves produced by the Monte  Carlo simulation of Hua (1997)  
fit quite well with values of $b \simeq 1$ and we have therefore adopted
this value for this parameter in the remaining of this work.

\subsection{The QPOs}

Given that the present model purports to provide a rather generic 
account of the variability of accreting sources we feel that, even at this 
early stage of its development, it should also be able to provide a 
generic account of the features occasionally occuring in the PSDs 
of these sources, known as Quasiperiodic Oscillations (QPOs). These features 
have attracted the attention of both theorists and observers, because
it was considered that their quasiperiodic nature would lead to clues
about the dynamics of accretion onto the compact object not available
in their largely aperiodic light curves. 

Since our present goal is not the study of the subject of QPO, 
we forgo any discussion, references or models associated
with this subject matter with the exception of the reviews by van der Klis 
(1995, 1998), wherein the interested reader can find more about the QPO 
phenomenology, systematics and additional references. 
However, we would like to demonstrate that the tenet of our model  - 
namely that features of PSD correspond to features in the electron 
density of the hot corona - can indeed provide an account of the QPO 
phenomenon and some of its systematics. 

In fig. 2a we present the response function of the following configuration:
Two concentric, uniform shells, each of $\tau_0 =1, ~ kT_e = 100$ keV, 
one extending from $r=0$  to $r = 1$ light second, while the second from 
$r_s = 10$ light seconds to $r_s + \Delta r$ with $\Delta r/r_s \ll 1$. 
Soft photons of energy
0.1 keV are released at the center, $r=0$, of the configuration. Each
panel of fig. 2a corresponds to the response function for photon in the
energy range denoted in the figure. One can distinguish a cut-off at
$t \simeq 1$ sec, indicative of the exponential drop-off of the photons 
escaping from the inner shell and a second one at $t \simeq 10$ sec due 
to escape through the outer thin shell. One should also note the 
presence of an additional peak at $t \simeq 20$ sec, corresponding to 
photons which were reflected by the outer shell (rather than transmitted 
through it), escaping at the opposite side of the configuration.
It is our contention that it is the presence of these photons which is
responsible for at least some of the aspects of the QPO phenomenon.

In fig. 2b we exhibit the power spectra corresponding to the shots 
given in fig. 2a. There is an apparent QPO peak at $\nu \simeq 1/20$ sec 
as well as harmonically spaced peaks, indicative of the power associated 
with the time scale $t_0 \simeq 2r_s/c$,  corresponding to the light 
crossing time of the outer shell. This additional power in the 
PSD is thus related to the spacial rather than timing properties of
an otherwise stationary configuration with generally stochastic 
soft photon injection. 
 As shown in figure 2, the QPO contribution 
increases with photon energy, a fact also in agreement with observation. 
This model provides a straightforward account of this effect: The 
larger the photon energy, the greater the number of scatterings it 
has undergone and therefore the greater the probability that is has crossed
the outer shell, thereby increasing the QPO contribution to the PSD. 
We view this dependence of the QPO fractional power on the photon 
energy as an interesting interplay between the spectral and timing
properties which the present model can easily accommodate within
its general framework.

The presence of multiple QPOs in the PSD of accreting sources implies 
therefore, within our model, the presence of multiple shells similar 
to that responsible for the PSD of figure 2. We do not know at this 
point the dynamics which would lead to the presence of such features;
the purpose of the present note is to simply indicate that their presence 
can lead to QPO-like features similar to those observed. Furhtermore, 
while features in the spacial electron distribution can indeed produce 
QPOs, one should be cautioned that they are not necessarily  the sole 
cause of all QPO features, and that more conventional models may be 
actually responsible for a number of them. We are in the process
of studying these possibilities in greater depth (Kazanas  \& Hua 
in preparation).

\section{THE LIGHT CURVES}

As discussed in the introduction, it is our opinion that one of the major 
obstacles in understanding the dynamics of accretion onto compact 
objects is the difficulty in providing concrete models of their observed  
time variability which could be easily compared to observation; in our
view, this lack of models is largely due to to the aperiodic character 
of their light curves, which offers few clues on the mechanism 
responsible for their formation. It is our contention that the light 
curves associated with the high energy emission of accreting compact 
sources is due, to a large extent, to the stochastic nature of Compton 
scattering in an optically thin, hot, extended medium, which is 
responsible for the formation of their spectra, coupled with a 
(probably) stochastic injection of soft photons to be Comptonized. 
Our proposal, therefore, is that the observed light curves consist 
of the incoherent superposition of elementary events, each triggered 
by the injection of a soft photon pulse into the extended corona of 
hot electrons discussed above. To simplify matters, we assume that 
the soft photon injection takes place at the center of the corona and
it is of vanishing duration; therefore 
the resulting high energy light curve should be the incoherent sum of 
pulses with shapes given by the corresponding response function 
discussed in the previous section.

Following this prescription, one can easily produce model light curves of the 
resulting high energy emission. These will have the form
$$F(t) = \sum_{i=1}^N Q_i g(t - \tau_i)  \eqno(6)$$

\noindent
The variable $\tau_i$ in the above equation is a random variable indicating
the injection times of the individual shots, while $Q_i$ is their normalization, 
a quantity which in our specific model depends on the number of soft photons
injected at each particular shot event. Clearly, one could in principle
arrange for any form of the PSD by choosing the values of the parameters 
$Q_i, \tau_i$ from appropriately defined distributions. While the values 
of $Q_i$ and $\tau_i$ may in fact be associated with certain distributions, 
we have no {\it a priori} knowledge of such distributions and 
they are in no way restricted by any compelling dynamical arguments. Therefore,
in order to avoid introducing 
extraneous information into our time series, we have chosen a constant 
value for all $Q_i$'s, while the values of $\tau_i$'s are chosen to be Poisson
distributed with a given constant rate. As such, the $\tau_i$'s are chosen
using the relation $\tau_i = - f \cdot t_0 \cdot log R_i$, where $R_i$ is a 
random number uniformly distributed between 0 and 1 and $f$ a real 
number, indicating the mean time between shots in terms of their rise time
$t_0$.

In Figures 3a and 3b we present two such model light curves. The parameters 
of the shots used in constructing these curves  were the same as those 
fitting the response function of figure 2a.  The two figures are different
in the value of the parameter $f$ i.e. the parameter which indicates the
mean arrival time between shots. The value of $f$ was set to $f = 3$ in 
figure 3a and $f = 10 $ in figure 3b. The two curves were produced with 
exactly the same sequence of random numbers, a fact which can be discerned
by identifying corresponding features in the two light curves. The random
character of the parameter $\tau_i$ leads then to an apparently incoherent
light curve in the sense discussed by Krolik, Done \& Madejski (1993),
i.e. of random absolute phases as a function of the Fourier frequency.

In figures 4a and 4b we present ``zoom-in" sections of the light curves  
of Figs 3a, 3b, corresponding to the same ordinal in the sequence of random 
shots. The shape of the shots becomes progressively asymmetric as the 
value of $f$ increases, since that gives time to the contribution of an 
individual shot to  the light curve to decrease substantially before 
the next one appears, thus manifesting the underlying asymmetry of the 
individual shots. At the same time, this also leads to an increase 
in the RMS amplitude of the variability. One should note that the model 
light curves presented in these figures consist exclusively of the 
superposition of shots which die out like power laws rather than 
exponentially in time, without the presence of a d.c. component. The 
brief rising part of the light curve lasts for a time interval of 
$\simeq \beta$ seconds, i.e. for the time it takes the earliest 
shot to die out, and it is a transient associated with the turn-on 
process. Beyond this point a steady -- state of rather well 
defined mean is established, which however is not particularly
smooth but characterized by large, aperiodic oscillations, 
the result of the random arrival and superposition of shots of shapes
similar to those of Fig. 1. 

It is apparent from the above that an RMS
value for the light curve can be established only for time scales $t 
\gsim \beta$. We thus propose that observations in which the RMS
values of the corresponding light curves depend on the observation 
interval indicate that these light curves have not been sampled for 
sufficiently long intervals i.e.  $t \simeq \beta$, even if the accretion
rate has remained constant during this interval. Conversely, the sampling
time over which a well defined RMS value of the light curve can be 
established, could be used to estimate of the time scale $\beta$ and thus 
the size of the high energy emitting source. This size could then be 
compared to that obtained through  study of the lags (see HKC) for consistency.

% additional text for light curve paper

The present model allows for several different ways of implementing a
change in the RMS amplitude of the model light curves. These are, in 
principle, related to observables and could be lead to provide novel 
insights concering the variability and dynamics of accreting sources. 
The simple prescription for generating the light curves given above
indicates clearly that increasing the value of $f$, i.e. the mean time between 
shots in units of their rise time, would result in larger RMS fluctuations of the 
light curves, should the rest attributes of the response function remain the
same. This particular case is discussed in more detail in \S 5.2 as it 
is expected to be correlated to the skewness of the light curves.

In this section we discuss the additional possibility of changes in the RMS
fluctuations of the light curves effected by changes in the value of 
the rise time of the shots $t_0$. For example, decreasing $t_0$ while keeping 
the  other parameters (size, temperature, density) of the hot corona constant, 
would lead to an increase in the source luminosity and a corresponding 
decrease in the RMS fluctuation amplitude without any additional changes in
the spectral or temporal properties of the source. 

However, if the time 
scales $t_0$ and $\beta$ are indeed related to a length scale associated 
with the size of the corona's inner and outer edges, this would generally 
depend on the macroscopic parameters associated with the accretion flow, 
most notably the accretion rate. 
To illustrate with an example the insights which can be gained by 
comparing such simple models to observation, we assume that the inner 
and outer radii of the corona $r_1, \; r_2$ (and for that matter the entire 
flow within the corona), scale proportionally to $v_{ff} \tau_{\rm cool}$,
where $v_{ff}$ is the free-fall velocity and $\tau_{\rm cool}$ the 
local cooling time scale. Assuming the latter to be inversely proportional 
to the local density, a prescription in accordance with, say, the models of
ADAF of 
Narayan \& Yi (1994), then all the length scales of the corona, expressed
in Schwarzschild radii, should be inversely proportional to the accretion rate 
of the flow, measured in units of its Eddington value. Therefore, an increase in 
$\dot m$ would lead not only to a decrease in $t_0$ and therefore in 
the RMS amplitude of fluctuations, but also in a concommitant decrease
of all other lengths of the system (e.g. its outer edge), resulting to an 
increase in the PSD frequencies corresponding to these features (e.g. 
the PSD low frequency break). 

Apparently such correlations between the source luminosity and the 
timing characteristics, while they may not be a universal phenomenon, 
have been observed in at least several sources. 
For example, van der Hooft et al. (1996) indicate that an increase in 
the luminostity of the Black Hole Candidate GRO J1719-24 leads to an 
increase in the QPO frequency at 0.04 Hz by a factor of $\sim 4$ while at 
the same time the RMS amplitude decreases by exactly the same 
amount preserving total variability power in the sense that the product
$\omega \vert F(\omega) \vert ^2 $ remains constant. These authors
indicate that scaling down all the frequencies of the PSD obtained when 
the source was in it high state, leads to a PSD indistinguishable from that
obtained when this source had a much lower luminosity. 
Similar general trends  have also been observed in the transient source
GRO J0422+32 (Grove et al. 1998) and are indicative that this behavior 
does not represent an isolated phenomenon associated with a particular
source. At the same time, the simple example discussed above indicates
how modeling the aperiodic variability of these sources could lead to 
new insights and probes of the dynamics of accretion onto compact 
objects.
                                                                                                                               
Concerning the morphology of the light curves given in figs. 3a and 3b,
eye inspection reveals shots with a variety of time scales, and indeed 
we believe that a distribution of such shots can be found, should one care 
to view and model them as such (see e.g. Focke \& Swank 1998). However, 
none of these additional time scales or shots have been used as input 
in this particular simulation. They result simply from the superposition
of a large number of shots with power law tails which extend to $\sim 1$
second. Zooming-in to the highest time resolution, one can indeed 
discern shots with rise times of the order of $t_0 \sim 10^{-3}$ sec
in agreement with
observations (Meekins et al. 1984). These shots are indeed the individual
elementary events (with the power-law extended tails) which comprise our 
model light curves. However, given the rather limited time range over 
which one can distinguish the contribution of such individual shots over 
the flux produced by the incoherent sum of their ensemble (or the 
detector statistics), one would most likely attempt to fit their shape 
with an exponential rather than a power law, because their long, extended 
tails cannot be discerned in the data. However, while fits in the time 
domain fail to uncover the true structure of these shots because of 
their crowding, the PSD can achieve this very easily and reliably. 

The above analysis makes clear that our model for the variability 
of accreting sources stands in stark contrast to other models put
forth to date to account for the ``flicker noise" character of the 
observed variability of BHC (Chen \& Taam 1995; Takeuchi, Mineshige 
\& Negoro 1995): Rather than attributing the observed PSD characteristics
entirely to variations in the mass flux  onto the compact object, it 
attributes them, for the most part, to the spacial distribution of 
electrons in the Comptonizing cloud and the stochastic nature of 
Compton scattering. The accretion rate is no doubt variable, 
especially at the shortest intervals - smallest radii (e.g. the soft 
photon injection events), however its variability is consistent with 
a constant ``averaged-out" accretion rate on longer time scales, as 
indicated by the coherence measurements and their interpretation (HKT). 
More importantly, this analysis provides a direct association between 
the physical properties of the scattering ``corona" and specific 
features of the PSD, in particular its low and high frequency breaks. 
Since the scattering properties of the extended corona (i.e. $p, ~T_e, 
~n_1, ~\tau, ~r_2$) affect both the spectral and the temporal properties 
of the emitted radiation in well defined fashion, this model implies the 
presence of certain well defined spectro-temporal correlations and 
provides the motivation to search for and model them in detail. 
Because such correlations will have to be related eventually to the dynamics 
of accretion, they could serve as a means for probing these dynamics
along the lines of the simple example given above.
Finally, the apparently random character of the observed light curves
(Lochner et al. 1991) is attributed to the ``random" (Poisson) injection 
of soft photons and the stochastic nature of the Comptonization process.

\section{THE TIME AND PHASE LAGS}

The fact that the variability associated with the light curves of figure
3 is due in part to Compton scattering, rather than, for instance, 
the modulation of the accretion rate, affords an additional probe of the 
properties of scattering medium, namely the study of time 
or phase lags in the light curves of two different 
energy bands as a function of the Fourier frequency. This particular 
issue has already been discussed in KHT, HKT  and in greater detail in
HKC. It bears on the fact that time lags between
photons of different energies depend on  the scattering time of the 
plasma within which the Compton scattering takes place. In a corona 
with a density profile given by Equation (1), there is a linear relation
between the scattering time and the crossing time of a given decade in
radius, while in addition, the probability of scattering within a given 
(logarithmic) range in radius is constant (for $p=1$). Consider now the 
light curves in two different energy bands: The escaping photons in the 
higher of two energy bands suffer, on average, a larger number of 
scatterings, which, for $p=1$, take place with equal probability at all 
radii; the information about the radii at which the additional scatterings 
took place is imprinted in the Fourier structure of these time lags. 
The linearity between the scattering and the crossing times then suggests 
that the lags grow proportionally to the Fourier period $P$.

We have repeated the lag analysis outlined in the previous references,
this time with model light curves appropriate for two different energies,
generated artificially using the algorithm described in the previous 
section. These were generated by sums of shots as demanded by Eq. (6), 
with the function $g(t)$ in each sum being the response function 
corresponding to a given photon energy. As noted in KHT, the very 
fact that higher energies require
longer residence times of the photons in the scattering cloud, leads to
small but significant (from the point of view of the lags; see HKC) 
differences in their corresponding response 
functions. In particular, the power law part of the response function
is slightly flatter (larger $\alpha$) and extends to slightly larger
times (larger $\beta$). While eye inspection cannot discern any difference
in the shape of the light curves corresponding to the two different 
energies, they are easily manifest in the Fourier decomposition of the
time lags. As pointed out in HKC
these differences in the individual shot profiles suffice to produce 
lags in general agreement with observation. 

Figure 5 presents the phase lags of our model light curves as a function 
of the Fourier frequency. Its magnitude
and Fourier frequency dependence is very similar to those corresponding
to galactic BHC analyzed by Miyamoto et al. (1991) and to that of
the X-ray transient source GRO J0422+32 (Grove et al. 1998). 
The corresponding time lags of the light curves so generated 
are obtained by simply dividing the phase lags by the Fourier frequency.
The fact that the phase lags are almost constant over the entire range
in Fourier frequency of figure 4 suggests that the corresponding time lags
will be roughly proportional to the Fourier period $P$, as discussed 
in HKC.

One could apply the arguments presented in the previous section on the
dependence of the PSD shape and features to infer the dependence
of the lags on the accretion rate; the contraction of all scales associated 
with the corona 
argued there should also reflect to a decrease of the corresponding lags 
with increasing accretion rate. As indicated by Cui et al. (1997) 
such a dependence has been observed in Cyg X-1 providing further 
evidence in favor of this specific model.

\section{THE AUTOCORRELATION AND TIME SKEWNESS FUNCTIONS}

In addition to the information provided by the Power Spectral Densities 
(PSD) and the phase or time lags discussed above, insights into the 
nature of variability of the BHC sources can also be obtained from  moments
of the light curves in the time rather than the frequency domain. These 
have been used in the analysis of the light curves of accreting compact 
objects several times in the past (Lochner et al 1991). Since we are able 
to produce models of these light curves we feel that it is instructive to 
compute the corresponding statistics associated with them so that one could 
directly compare them to those associated with the light curves obtained 
from  observation. At present we will pay particular attention to two such 
statistics, the autocorrelation function and the time skewness of the
light curves.

\subsection{The Autocorrelation Function}

This statistic provides a measure of the dependence of the flux (counting 
rate), $F(t)$, of the source at a given time $t$ on its flux, $F(t+\tau)$, 
at a prior time. Assuming
that the source variability consists of flares of a particular time
scale, the autocorrelation function provides a measure of this time scale.
While the information contained in the autocorrelation
function is related to that of the PSD through the fluctuation -
dissipation theorem, since this statistic is also used to gauge the 
variability of accreting sources, for purposes of comparison, it is 
instructive to provide its form for the model light curves we produce 
in conjunction with the PSD.

Given that our model light curves are the incoherent 
sum of a large number of shots of the form given by Eq.(4), in 
order to exhibit directly the effect of the random injection of
shots used to produce the light curve, we have 
chosen to compute the autocorrelation function in two ways: (a) through
the convolution of the response function  $g(t)$ with itself, i.e. 
$$ACF(\tau) = \int_0^{\infty}  g(t)   g(t + \tau) \; dt  \eqno(7)$$

\noindent where $\tau$ is the associated time lag. 
(b) directly from the model light curves produced by the procedure 
described above. Considering that the light curve consists of measurements
of the flux $F$ at $N$ points in time, denoted as $t_i$, separated in time
by an interval $\Delta \tau$, the autocorrelation function at a 
given lag, $\tau = u \cdot \Delta \tau$, is given by the sum (ignoring
normalization factors)
$$ACF(\tau ) = \sum_i^{N-u} [F(t_i) - \bar F]
[F(t_i+ u\cdot \Delta\tau) - \bar F]     \eqno(8)$$
\noindent where $\bar F$ is the mean value of the flux over the 
interval we consider. 

The results of these two procedures are shown in figures 5a, 5b. As it 
can be seen, the autocorrelation function computed by these two 
procedures are consistent with each other, with the fluctuations 
at the largest lags of fig. 5b due to the statistical nature of our light 
curve. This latter one is also very similar to that computed from the
Cyg X-1 light curve by Meekins et al. (1984) and Lochner et al. (1991). The 
similarity of the autocorrelation function  of that obtained from the 
observations further corroborates the model we have just presented.

At this point we would like to note that, usually, the data associated
with the autocorrelation function are presented, (e.g. Meekins et al. 1994;
Lochner et al. 1991), in linear time coordinate; 
we believe that this presentation masks most of the interesting physics 
which are contained in the interval near zero. It is our contention that
in the study of systems whose variability apparently spans several decades
in frequency, like the accreting compact objects considered 
in the present note, the use of logarithmic rather than linear coordinates
is instrumental; it is only in terms of former that one can capture the
entire range of the physical processes involved. The form of the 
autocorrelation  functions corresponding to those of Figs. 5a, 5b in 
logarithmic time coordinate are given in Figs. 5c, 5d.

\subsection{The Time Skewness Function}

This statistic measures the time asymmetry of a given light curve, i.e. 
whether the latter is composed of pulses having sharper rise than 
decay or vice versa. Because the autocorrelation function is symmetric
in the lag variable $\tau$, it cannot give any such information about 
the shape of the light curve. This property can be assessed by computing
moments of the light curve higher than the second.  In particular, 
the third moment, $Q(\tau) ~(\tau= u\cdot \Delta \tau)$, as defined 
in Priedhorsky 
et al. (1979), provides  the proper statistic. For a light curve given as an
array of the flux $F(t_i)$ as described in the previous subsection, 
the skewness is given by the sum (ignoring again the normalization factors)
$$Q(\tau) = \sum_i^{N-u}[F(t_i) - \bar F][F(t_i+ u\cdot \Delta
\tau) - \bar F][F(t_i) - F(t_i+ u\cdot \Delta\tau)]       \eqno(9)$$

Given that our model light curves are sums of shots with a unique time
profile, one can  infer {\em a priori} several properties of the resulting
light curves: The form of the shot profile (Eq. 4) suggests that, 
since the shots are asymmetric in time, the corresponding light curves 
should be also asymmetric, at least in situations in which the 
contribution of individual shots can be perceived. 
However, the presence of the long power--law tails associated
with the individual shots, suggests that  over sufficiently long time 
scales, which encompass a large number of shots, the light curves 
should be largely symmetric in time. This argument
is born out by both inspection of our model light curves and computation
of the skewness parameter.

The prescription for creating model light curves discussed in section 3, 
offers the possibility of testing the above arguments by producing 
light curves with the proper characteristics, through variation of 
one or more of their control parameters.  The specific parameter in 
this case is the mean time between shots $f$. In figure 6 we present the
skewness function of light curves corresponding to two different values 
of this parameter, namely $f = 2, 10$. As expected, for the small values 
of this parameter the light curves are indeed symmetric as it can be 
assessed  both by inspection and the value of skewness. For $f=10$ 
the light curves become distinctly asymmetric out to time lags roughly 
$10 t_0$, beyond which they appear again symmetric due to the superposition 
of a large number of shots. 

This specific property then offers itself to observational testing: 
One should note that large values of $f$ lead not only to non-zero 
short time skewness, but also to large RMS fluctuations of the light 
curve as alluded in \S 3; in fact, the larger the $f$-value,  
the larger the corresponding 
RMS fluctuations and also the value of the lag $\tau$ for which the 
skewness, $Q(\tau)$, of the light curve deviates from a non-zero value. 
To the best of our knowledge, such a correlation has never been 
proposed or sought in the data. It would be of interest to see to what 
extent it is born by observations. 

\section{CONCLUSIONS, DISCUSSION}

We have presented above a general prescription for producing model 
light curves of accreting compact objects. Within our model,
the observed {\em aperiodic} variability of these light curves 
is due to the stochastic nature of the Comptonization process in 
conjunction with the soft photon injection near the compact object by a
Poisson process. Our prescription thus accounts naturally for the apparent 
lack of coherence in the absolute phase of the observed 
light curves and the apparent high coherence in the relative phase
of the light curves of two different energy bands, since both the hard and
soft shots have the same origin in the impulsive injection of the soft
photons. Concerning the most common test of variability, namely the PSD,
our model relates it
to spacial rather than timing properties of these systems, in particular
to the spacial distribution of electrons in the Comptonizing hot corona.
Thus it produces PSDs in agreement with observation and provides a novel
framework within which one can easily accommodate the existence of QPOs
and some of their systematics and dependence on the sources' luminosity.
Within the present framework for the variability of accreting sources, 
their timing and spectral properties are intimately related in a way 
that could allow one to probe of the dynamics of accretion onto the 
compact object through the use a combined spectro-temporal analysis. 
Finally, the model light curves produced using the prescription indicated
above have a morphology is in general agreement with the observed 
{\em aperiodic} variability of this class of sources. 

The morphology of the light curves has been examined by computation
of two statistical properties in the time domain, namely the autocorrelation 
function and the time skewness. These, as computed for our model light
curves, appear to be in good agreement with their (albeit limited) 
published literature forms corresponding to the light curves of the 
BHC Cyg X-1. Considering the simplicity of our models (a single type 
of shot, Poisson distribution in injection times), we are quite surprised that 
they work as well as they do. Our investigation points, in addition, 
to a correlation between the RMS variability and the skewness of the 
corresponding light curves, which we feel should be tested against the 
observations.

The models presented herein draw heavily on the ideas presented in KHT, 
HKT and HKC, namely of very extended ($\sim 
10^3-10^4 ~{\rm R}_S$) hot electron coronae with power law dependence 
of the electron density in radius. In our view, the importance of these 
models lies in the implied direct association between features in the 
observed PSDs and time lags (i.e. features of their Fourier domain
characteristics) with features in the spacial domain (i.e size, radial 
density structure). Such an association is not dictated in any of the, 
albeit very few, alternative models of BHC variability, and to the 
best of our knowledge, neither has been proposed before in the literature. 

It is of interest that both the sources' sizes and density structures, as
deduced from our models (see also KHT, HKT, Hua et al. 1997), are 
incommensurate with those predicted by the most popular models.
These models generally require the sources' size to be only a few R$_S$
(Shapiro, Lightman \& Eardley 1976), leading to very narrow range in 
density (which in the present framework can be considered uniform) and 
therefore to an equally narrow range in time lags, in disagreement with 
observation.  The recent, popular ADAF (Narayan \& Yi 1994) are 
indeed extended in radius, as demanded by our models,  but their 
density profiles are proportional to $r^{-3/2}$, rather than the 
$r^{-1}$ profiles preferred by our fits to most (but not all) of the 
time lag observations obtained to date (HKC), suggesting that a variant 
of these models maybe more appropriate in describing the dynamics of
accretion in these sources. It is nonetheless
important to  point out that both of these density  profiles 
have been associated with data from the same source (Cyg X-1), 
at different viewing periods, indicating that, according to the 
present model,  the structure of a given source can vary drastically 
as a function  of time. 

We believe that the present model is sufficiently well defined and 
makes concrete enough predictions to allow its falsification or 
confirmation by 
more detailed observations. As such, we expect observations in the
time domain to be of vital importance.
Because the timing and spectral properties are intimately related within
our model, testing the particular paradigm would most
likely involve a combination of spectral and temporal correlations. 
We have provided fits of our models to a small set of observations,
and derived the corresponding physical parameters of the scattering
coronae for these systems; these indicate a great departure from 
our previous notions as to what they should be. 
We do not know as yet how to justify the values of the parameters 
obtained by our models, however this is not the purpose of the 
present paper. We simply hope that these models will stimulate additional
observational scrutiny and re-analysis of the data within their 
framework,  leading possibly to novel insights which will further our
understanding of the dynamics of accretion onto the compact object. 

Last but not least, these models will have to be modified to 
incorporate additional spectral features such as the reflection
features and the Fe lines, which the more conventinal models 
of thin, cold disks and their associated hot coronae have addressed
so far with significant success.

\begin{acknowledgments}
We would like to thank W. Focke, C. Shrader and W. Zhang for a number of 
interesting and informative discussions on the variability of accreting
black holes.
\end{acknowledgments}

\vfill\eject

\centerline{\bf Figure Captions}
Figure 1. 
The response function a corona with $p=1$ and  $r_1 = 6.35 \times 
10^{-3}$  light sec $=1.9 \times 10^8$ cm, $r_2= 6.24$ light sec$ = 1.87 
\times 10^{11}$ cm, $T_e = 100$ keV, $\tau_0 = 1$ and $n_1 = 10^{15}$ 
cm$^{-3}$ (long dashed curve) along with its fit by Eq. (4) with  
$\alpha = 0.4, ~\gamma = 1.5$, $\beta = 10$ sec, $t_0 = 0.02$ sec, $b=1$,
(solid curve) and the corresponding power spectrum (short dashed curve).

Figure 2. 
The response functions (2a) and the corresponding power spectra (2b) 
associated with the configuration consisting of a uniform sphere 
of $\tau_0 =1$ and radius $r = 1$ light second, surrounded by a 
thin shell of the same Thomson depth $\tau_s =1$ and $r = 10$ light
seconds.

Figure 3.
Model light curves constructed using the prescription of Eq. (6). Both 
curves use $\alpha = 0.5, ~\gamma = 1.5$, $\beta = 1.5$ sec, $t_0 = 0.001$ 
sec, $b=1$ but two different values of the mean arrival time between 
shots. The corresponding parameter $f$ takes the values $f=3$ (Fig. 3a) and
$f = 10$ (Fig. 3b).

Figure 4. 
``Zoomed-in" sections of the light curves of Figs. 3a and 3b respectively.
The two sections correspond to the same ordinal of the random shots 
at the beginning of each figure.

Figure 5. 
The phase lags corresponding to two sets of light curves generated as 
described in the text. The light curves had the following parameters
(a) $\alpha =0.5$, $\beta = 16$ sec, $t_0 = 0.02$ sec. (b) $\alpha =0.55$,
$\beta = 16$ sec, $t_0 = 0.02$ sec. (c) $\alpha =0.6$, $\beta = 16$ sec, 
$t_0 = 0.02$ sec. The two curves correspond to the lags between: (a) - (b)
(solid line) and (a) - (c) (dotted line).

Figure 6.
(a) The autocorrelation function computed using the 
corona response function $g(t)$ (Eq. 7) with $\alpha = 0.38$, 
$\beta = 10$ sec and $t_0 = 0.02$ sec. (b) The autocorrelation 
function  computed from a model light curve, generated according
to Eq. (6) with $g(t)$ as above. (c) Same as in figure 6a plotted
in logarithmic time coordinate. (d) Same as in figure 6b plotted 
in logarithmic time coordinate.

Figure 7. 
The skewness function as computed for two model light curves generated using
Eq. (6) with $g(t)$ parameters equal to those of figure 6 and two different 
values of $f$, $f=2$ (a) and $f=10$ (b). The change in the sign of skewness
with $f$ at short time scales is apparent in the figure.

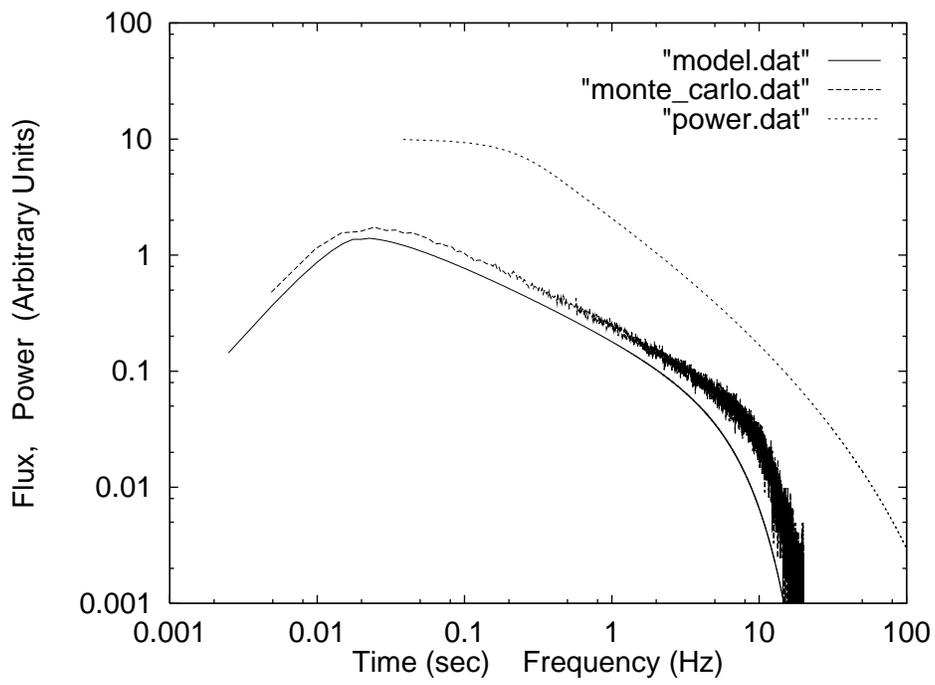
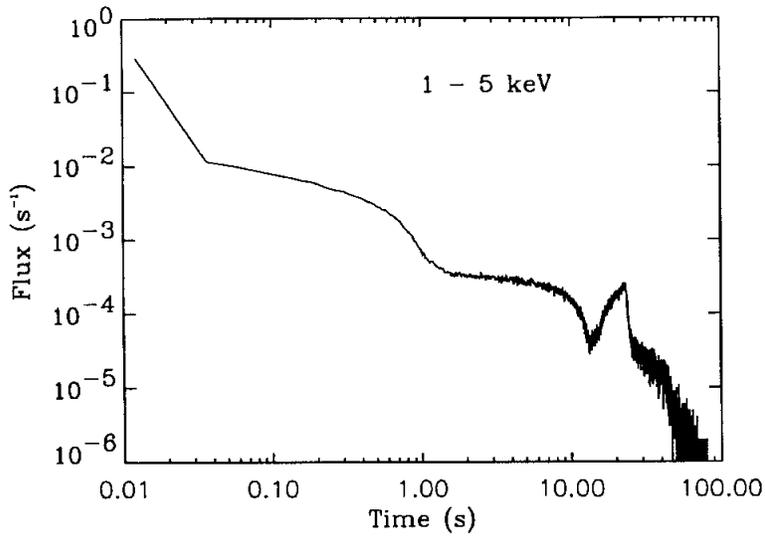
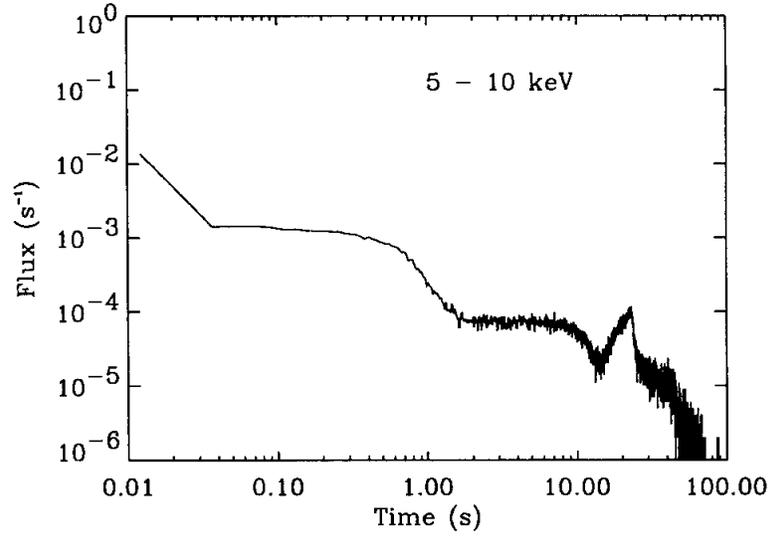
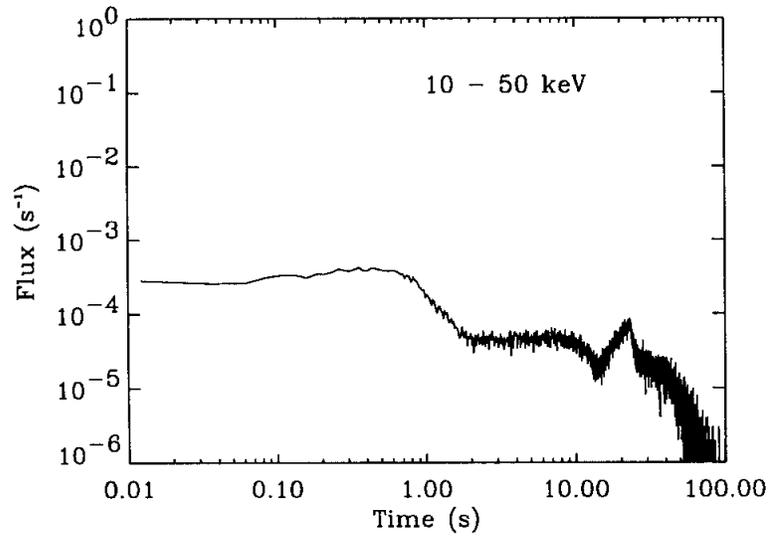
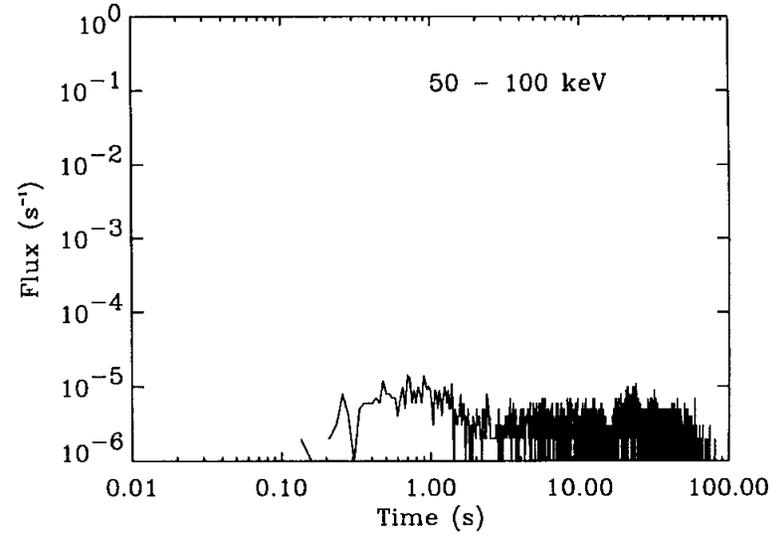
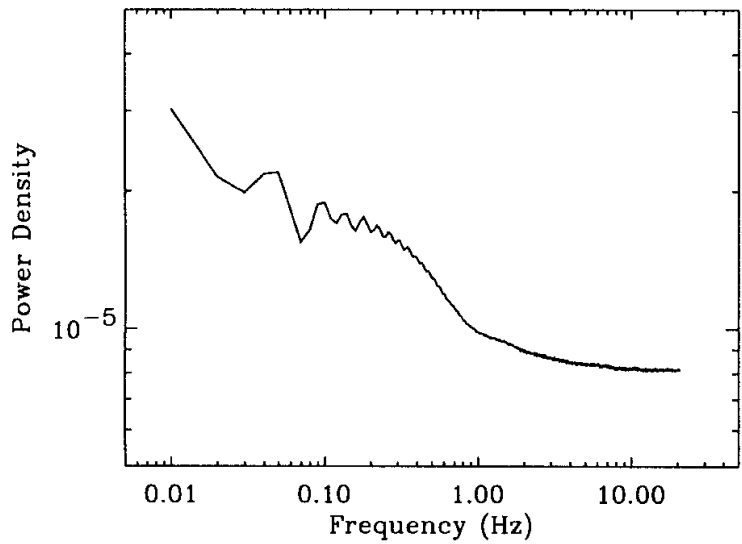
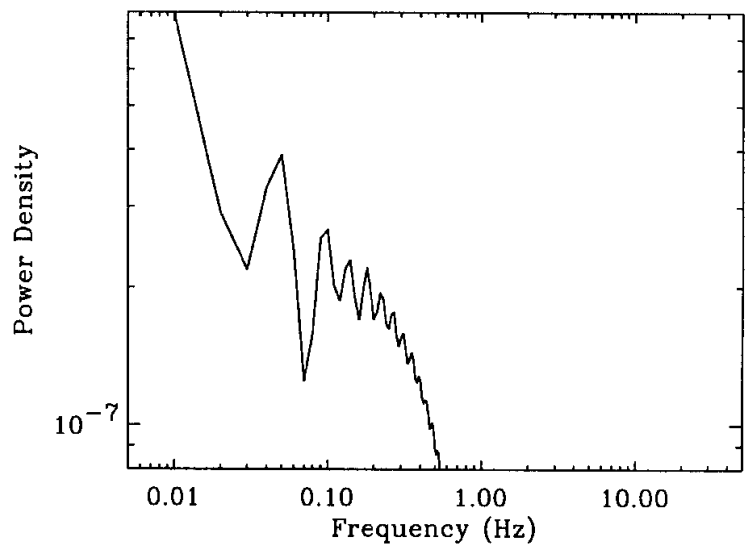
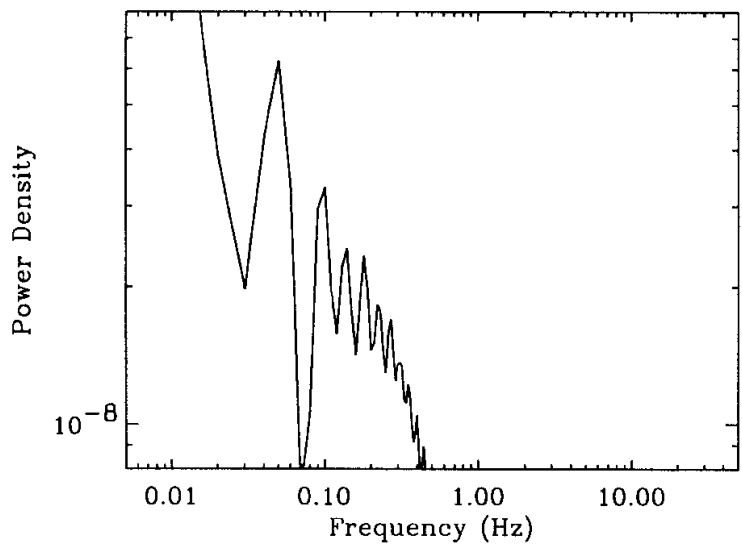
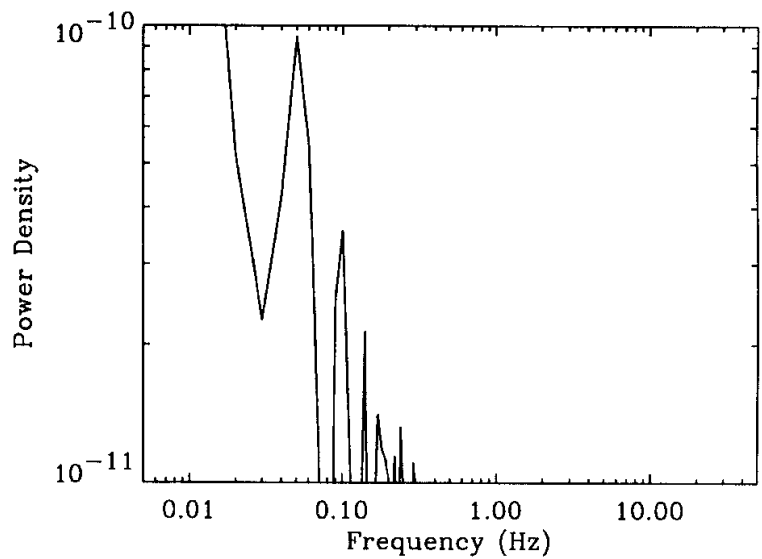
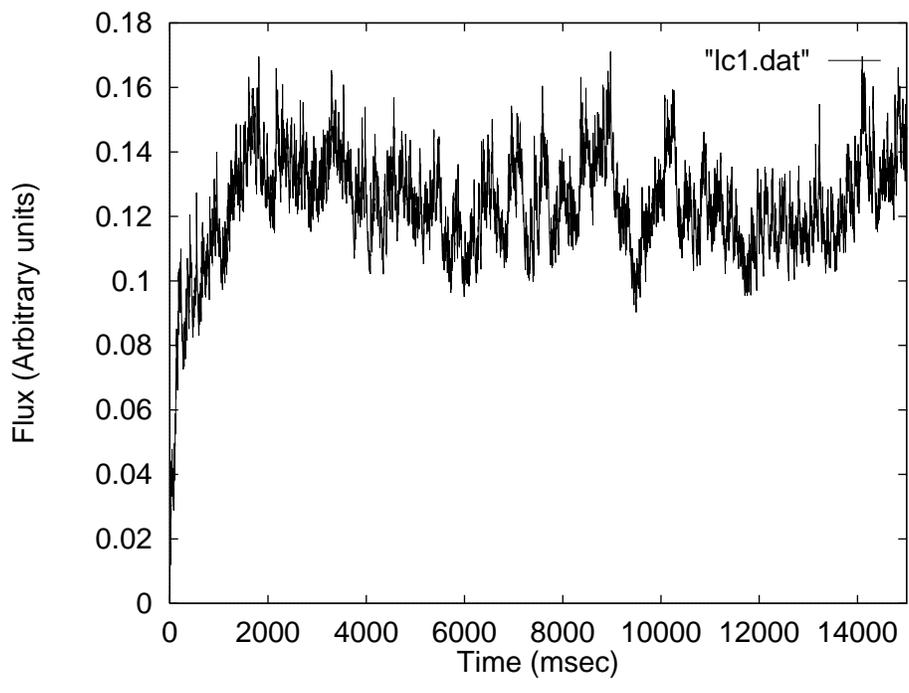
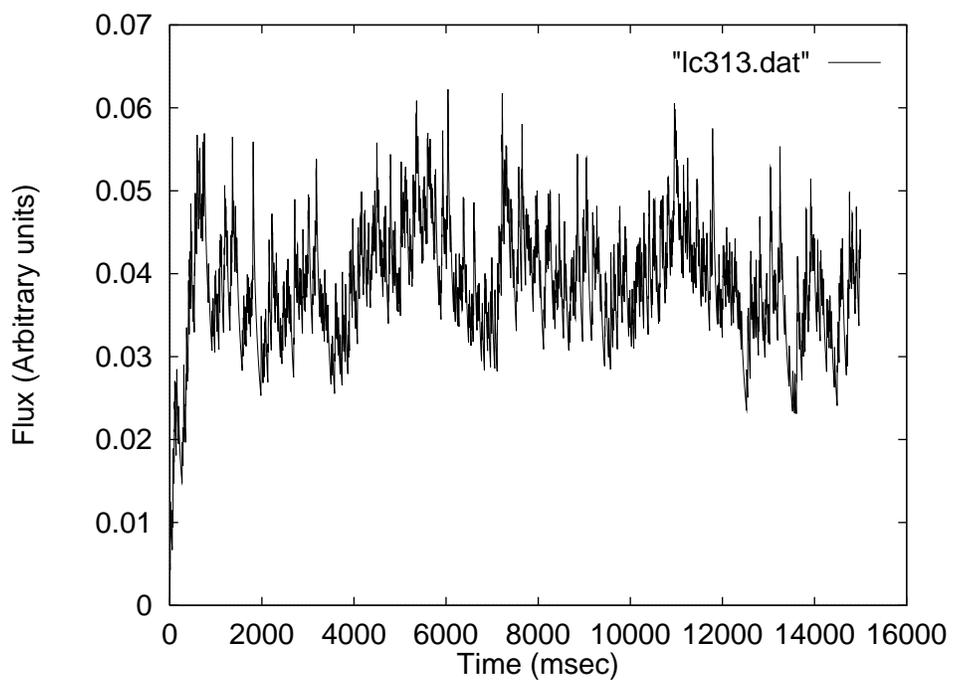
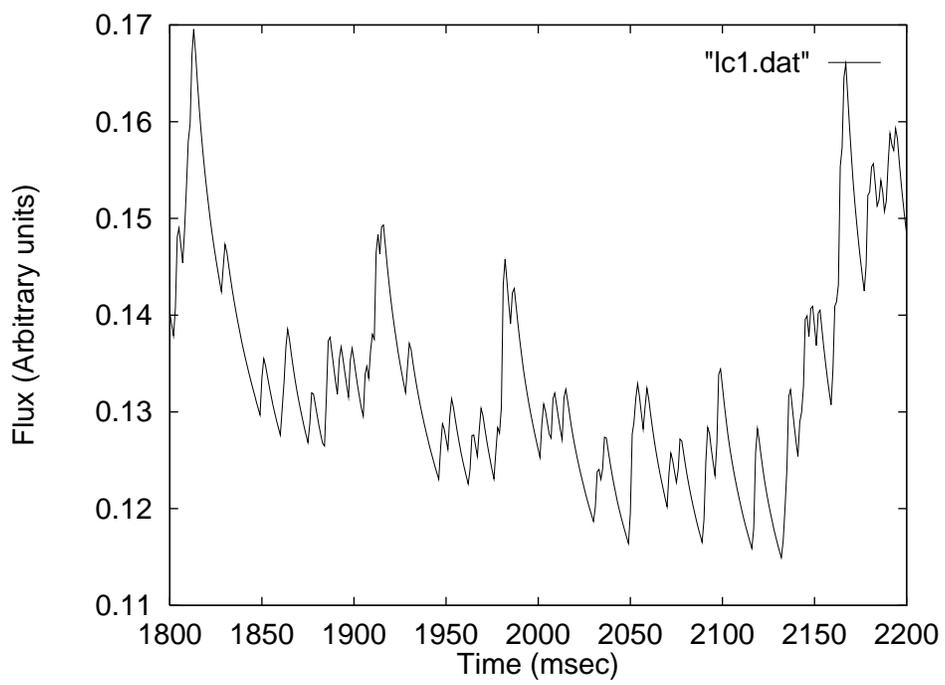
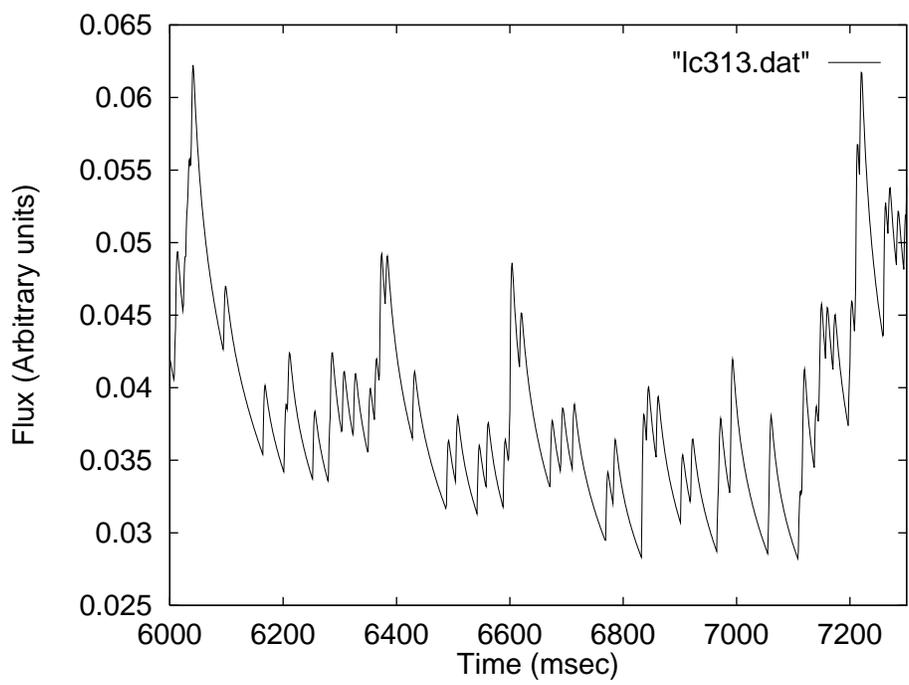
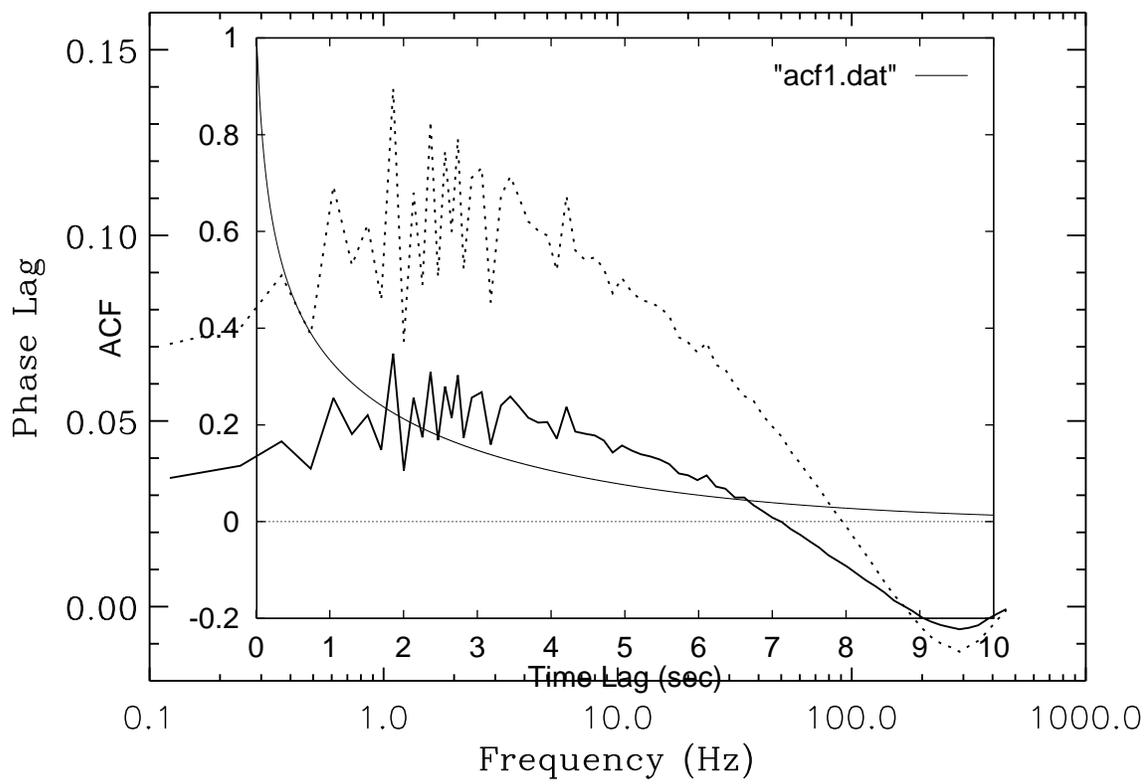
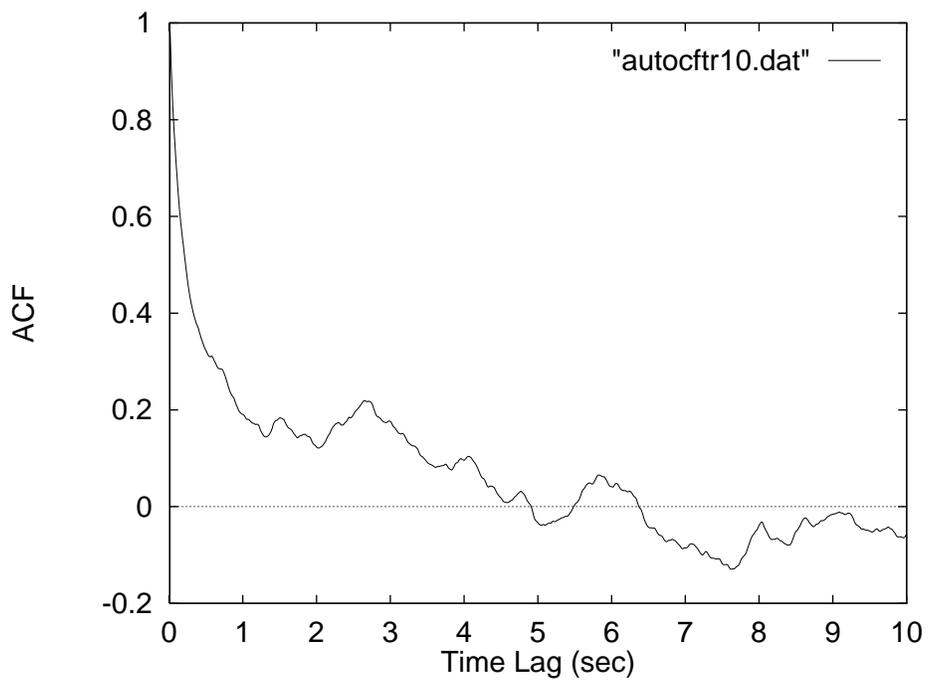
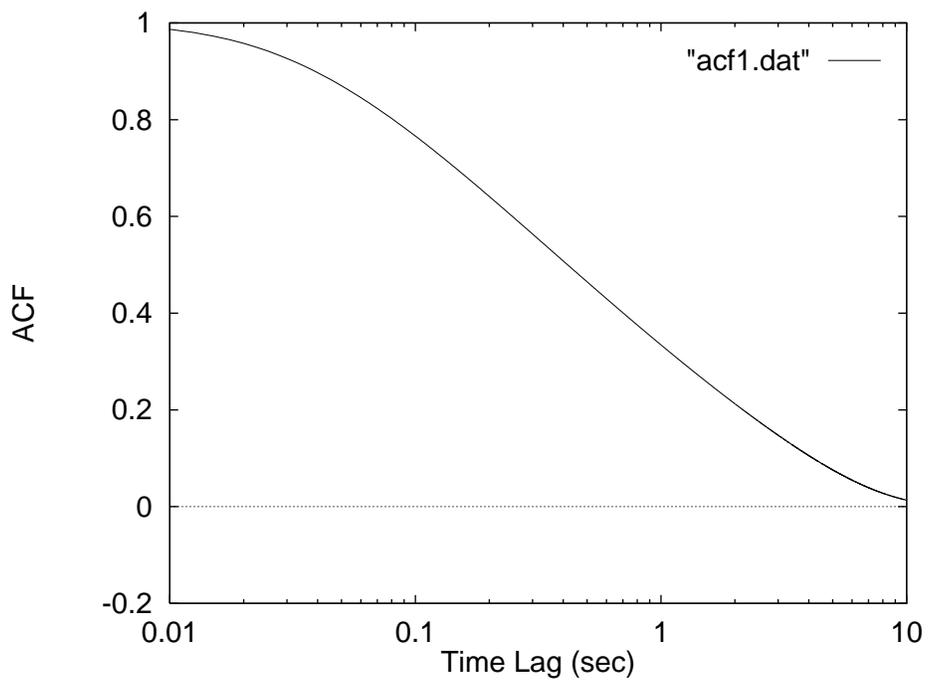
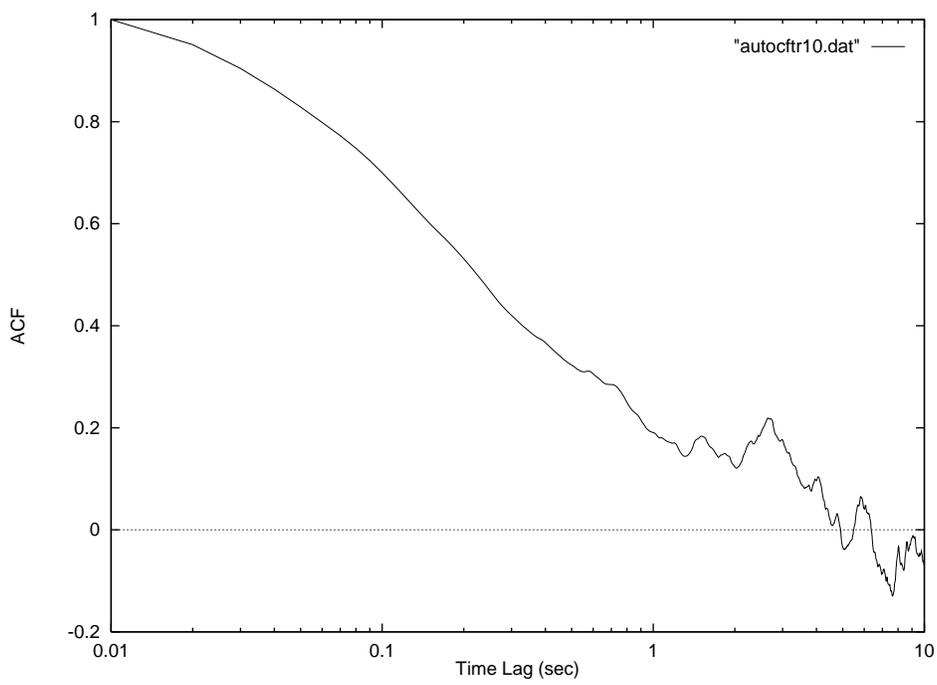
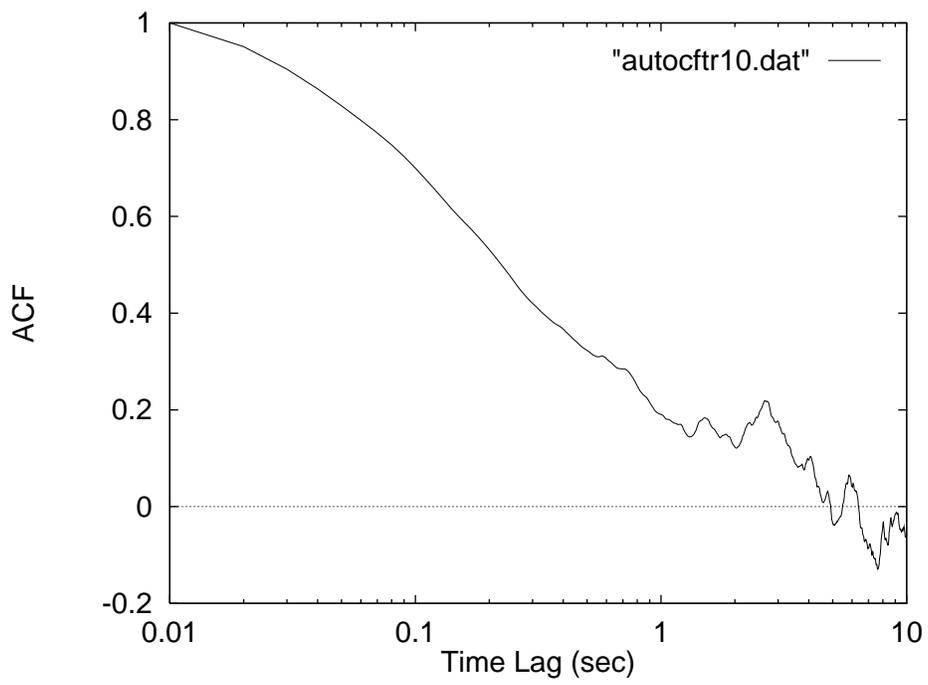
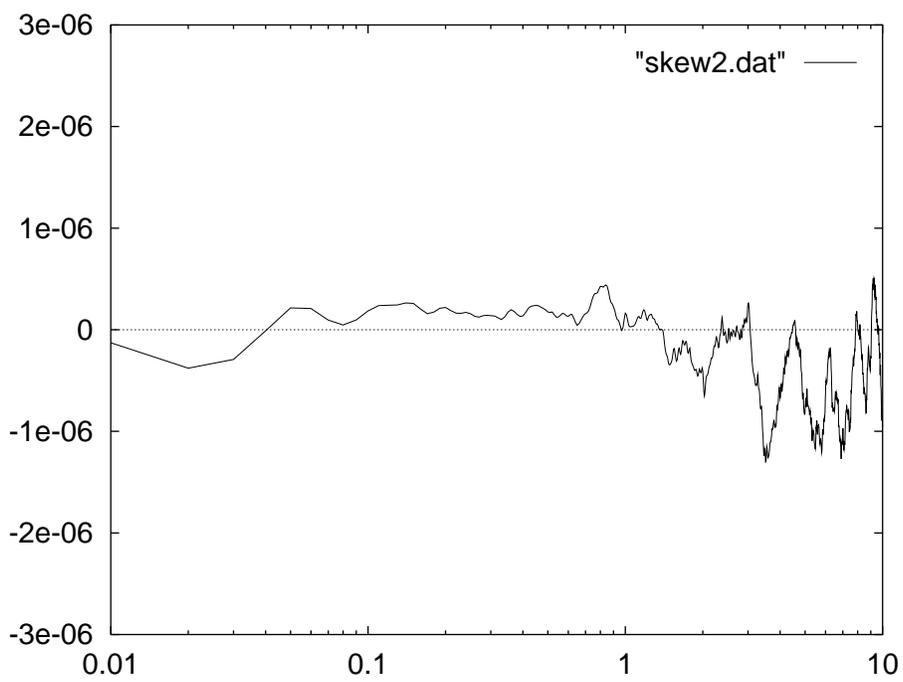
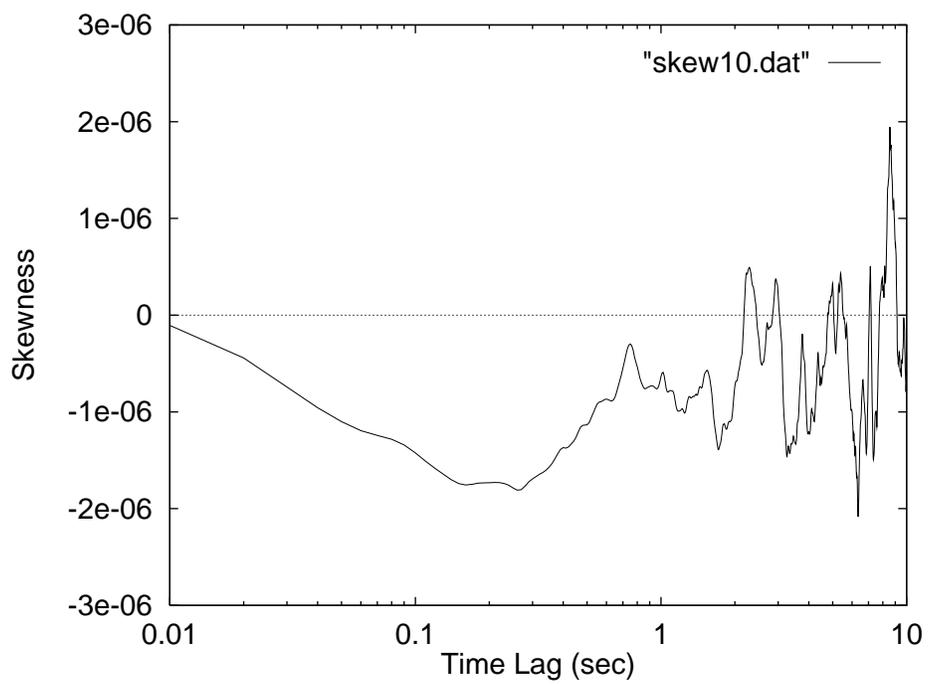
\end{document}